# OSPF extension for the support of multi-area networks with arbitrary topologies

Rui Valadas

*Abstract*—OSPF currently supports multi-area networking with two severe limitations, due to the distance vector approach taken in the inter-area routing protocol: (i) the multi-area topology is restricted to a two-level hierarchy, and (ii) globally optimal routing may not be achieved. In this letter, we propose an OSPF extension that overcomes these limitations by adopting a link state inter-area routing protocol. The extension applies to both OSPFv2 (IPv4) and OSPFv3 (IPv6), and is fully transparent to area-internal routers. Despite its simplicity, this extension may have a large impact in the operation of the current Internet.

## I. INTRODUCTION

OSPF [1] and IS-IS [2] are the main players in today's Internet intra-domain routing. They are implemented by virtually all vendors of routing equipment and have been widely deployed in most large IP networks worldwide. They belong to the class of *Link State Routing* (LSR) protocols. The present versions of OSPF are OSPFv2 (for IPv4) [1] and OSPFv3 (for IPv6) [3].

An OSPF network can be structured in multiple areas, a feature that eases the management of large networks and alleviates the memory requirements of routers. However, OSPF introduced several restrictions that (i) constrain the topology of multi-area networks and (ii) prevent globally optimal routing. This letter proposes an extension to OSPF that overcomes these limitations, and only requires modifications to the area border routers. In section II, we describe the current OSPF solution for multi-area networking, and discuss its limitations. In section III, we present the OSPF multi-area extension. Finally, in section III, we conclude the letter.

## II. OSPF HIERARCHICAL ROUTING

In OSPF the routers build and maintain a Link State Database (LSDB) containing the *topological* and *addressing* information that describes the network. The topological information corresponds to the network map (or graph), describing all routers and links between routers, and the addressing information corresponds to the address prefixes

Rui Valadas is a Full Professor at Dept. of Electrical and Computer Engineering, Instituto Superior Técnico, Universidade de Lisboa, and a senior researcher at Instituto de Telecomunicações, Portugal (e-mail: rui.valadas@tecnico.ulisboa.pt).

(IPv4 or IPv6) assigned to routers and links. OSPF includes several mechanisms to keep the LSDB updated at all routers, namely the Hello protocol, the reliable flooding procedure, and initial LSDB synchronization process [1]. The Hello protocol detects the active neighbors of a router, the reliable flooding procedure disseminates the routing information originated by one router, and the initial LSDB synchronization process allows a fast update of the LSDB when a router joins the network.

OSPF is an intra-domain routing protocol, i.e. it is used inside *Autonomous Systems* (ASes). The information on address prefixes external to an AS, i.e. the *AS-external prefixes*, is injected into the AS through *Autonomous Systems Border Routers* (ASBRs). When an AS is large, containing dozens of nodes and links, the LSDB also becomes large, and some nodes may lack memory resources to store it completely. One way to overcome this problem is to structure the network in smaller *areas*, such that nodes only need to keep the network map of the area they belong to; significant memory savings can be achieved in this way.

In a multi-area network using LSR, there is one LSDB per area, containing the network map of the area, and the addressing information of the AS. An *area-internal router* is unaware of the network topology of other areas. However, it still needs to get information on the destinations available outside its area. This information is obtained through an *inter-area routing protocol* running among the *Area Border Routers* (ABRs), i.e. the routers located in the frontier between areas. ABRs keep as many LSDBs as areas they directly attach to. In OSPF, two types of destinations are advertised across the inter-area routing protocol: the *area-external prefixes*, i.e. the address prefixes that are external to an area but internal to the AS, and the location of *area-external ASBRs*. In multi-area networks, the routing information may be disseminated throughout the whole AS, i.e. with *AS-flooding scope*, or only within an area, i.e. with *area-flooding scope*.

Figure 1 shows a generic a multi-area network. The network has 8 routers and is structured in 4 areas; routers R2 to R6 are ABRs, routers R1 and R7 are area-internal routers (from the perspective of areas 1 and 4, respectively), and router R8 is an ASBR. In this example, the AS has three address prefixes that need to be advertised: ap1 assigned to R1, ap2 assigned to R7, and ap3 injected in the AS by R8. The figure includes costs assigned to each link, which are used to determine the shortest paths between network elements.

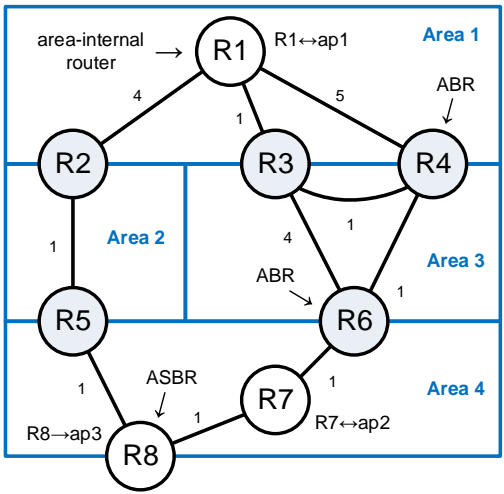

Figure 1: Multi-area network.

In OSPF, the inter-area routing protocol uses a Distance Vector Routing (DVR) approach. In this case, ABRs exchange *distance vectors* with their neighboring ABRs, to compute the shortest path cost and the next-hop ABR to each destination, using the usual DVR rules [4]. The distance vectors are (destination, cost) pairs, where the first element is the destination identifier and the second is the estimate of the shortest path cost from the sending ABR to the destination.

OSPF places two types of limitations regarding multi-area networking, due to the well-known convergence problems of distance vectors protocols [4]. First, the area design is constrained to a 2-level hierarchical structure. An OSPF multi-area network comprises a single upper-area, called *backbone*, and several lower-level areas that connect directly to the backbone. Lower-level areas cannot connect directly to each other and, therefore, all traffic between lower-level areas is forced to cross the backbone. This is why OSPF specifications refers to this type of networks as hierarchical networks, and not as multi-area networks. For example, the network of Figure 1 could not be implemented in OSPF, since the multi-area topology contains a cycle. OSPF includes an exception to this constraint through *virtual links*. Virtual links are tunnels allowing the connection to the backbone of areas not physically attached to it. However, virtual links require manual configuration and inherit the problems of static routes (e.g. no resilience to failures).

Second, OSPF imposes two restrictions on the way distance vectors are advertised: ABRs cannot advertise inside an area (i) routes to destinations internal to that area and (ii) routes to external destinations that cross that area. These restrictions may prevent globally optimal routing, i.e. the path selected between two nodes may not always be the shortest one. To see this, consider how R6 determines the shortest path cost and next-hop neighbor to ap1, in the multi-area network of Figure 1; we concentrate on the distance vectors sent by R3 and R4, to keep the explanation brief. R3 and R4 first broadcast to their neighboring ABRs the distance vectors computed from the LSDB of area 1; specifically, R3 sends vector (ap1, 1) to neighbors R2, R4, and R6, and R4 sends vector (ap1, 5) to R2, R3 and R6. When R6 receives these vectors, it determines that the shortest path cost to ap1 is 5 (via R3), which is still incorrect; the correct shortest path cost is 3 (via R4). To compute the correct information, R6 needs to receive vector (ap1,2) from R4, which R4 computes based on vector (ap1,1) sent initially by R3. However, due to restriction (ii), R4 cannot inject vector (ap1,2) on area 3, since the underlying route crosses this area (it is via R3). Thus, R6 keeps believing that the shortest path from itself to ap1 is via R3 with a cost of 5, which not optimal.

The OSPF LSDB is divided in records, named *Link State Advertisements* (LSAs), each describing a specific network characteristic. Each LSA has an *originating router*, i.e. a router responsible for its creation, updating, deletion, and dissemination. LSAs are disseminated independently of each other. In OSPFv3, the area topology is described by Router-LSAs and Network-LSAs, the area-internal and area-external prefixes are described by Intra-Area-Prefix-LSAs and Inter-Area-Prefix-LSAs, respectively, the domain-external prefixes are described by AS-External-LSAs, and the locations of area-external ASBRs are described by Inter-Area-Router-LSAs. OSPFv2 differs only in the way the area-internal addressing information is handled: the topological and addressing information is provided simultaneously by Router-LSAs and Network-LSAs, and there is no equivalent to the Intra-Area-Prefix-LSAs. All other LSAs have a direct equivalent in OSPFv2: Inter-Area-Prefix-LSAs are equivalent to Network-Summary-LSAs, and Inter-Area-Router-LSAs to ASBR-Summary-LSAs; AS-External-LSAs kept the designation of OSPFv2. The differences between OSPFv2 and OSPFv3 relate only to area-internal addressing information, and have no impact in the inter-area routing protocol. Excepting the AS-External-LSA, which has AS-flooding scope, all other LSAs have area-flooding scope. The LSAs that are disseminated through the inter-area routing protocol, i.e. the distance vectors, are the inter-area LSAs of OSPFv3 (Inter-Area-Prefix-LSA and Inter-Area-Router-LSA), and the summary LSAs of OSPFv2 (Network-Summary-LSA and ASBR-Summary-LSA). The information regarding an AS-external prefix is disseminated inside the AS using two LSAs: the AS-External-LSA advertises the actual prefix, and the Inter-Area-Router-LSA (OSPFv3) or the ASBR-Summary-LSA (OSPFv2) advertise the location of ASBR that injected the prefix.

## III. THE OSPF MULTI-AREA EXTENSION

In a multi-area network, the set of ABRs and connections between them forms a routing *overlay*, i.e. a logical network over the physical network utilized for the exchange of inter-area routing information. The graph representation of this overlay is a key element of our proposal. In this graph, nodes correspond to ABRs, arcs to intra-area shortest paths between neighboring ABRs, and arc weights correspond to their costs. The area-internal routers are not part of the graph. Moreover, each ABR is labelled with (destination, cost) pairs, where destination is either an address prefix or an ASBR available inside the areas it directly attaches to, and cost is the corresponding intra-area shortest path cost. Note that the various intra-area shortest path costs of the graph are obtained from the area LSDBs available at each ABR. Figure 2.a shows the overlay graph corresponding to the network Figure 1. Notice that the arc between R3 and R6 is labelled with weight 2, corresponding to the shortest path cost between these two routers, which is via R4.

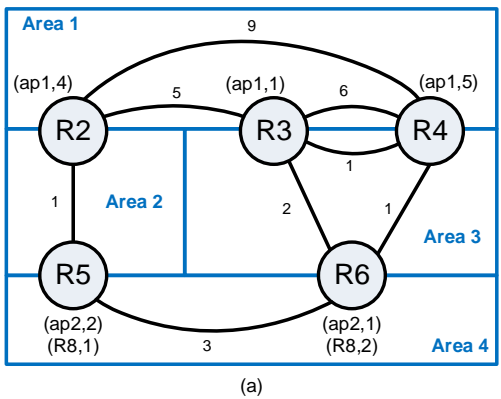

Figure 2: (a) Graph and (b) LSAs describing the ABR overlay of Figure 1.

The graph of ABR overlay immediately suggests an LSR approach to inter-area routing which, as it will be seen soon, overcomes the limitations of hierarchical OSPF described in previous section. In this approach, each ABR builds and maintains the graph of ABR overlay based on the contributions of all ABRs. Specifically, each ABR determines its local view of the ABR overlay, i.e. who its neighboring ABRs are and what are the costs of intra-area shortest paths towards them, using the LSDBs of its directly attached areas; it also determines the address prefixes and ASBRs available at these areas. This information is then disseminated with AS-flooding scope to all other ABRs. For example, router R6 disseminates information that is has neighbors R3, R4, and R5, reachable at costs 2, 1 and 3, respectively, and that it has direct access to address prefixes ap2 at cost 1, and to ASBR R8 at cost 2. When an ABR receives the local views of all other ABRs, it builds the overlay graph, and determines the shortest paths from itself to all network destinations. This approach replicates, at the ABR level, the way the network map is built inside areas, and is currently not supported by OSPF or any other LSR technology.

To describe the ABR overlay we introduce three new LSAs, named ABR-LSA, Prefix-LSA, and ASBR-LSA. These LSAs are disseminated with AS-flooding scope, and are only originated and processed by the ABRs; they will be referred to as *overlay LSAs*. Figure 2.b shows the LSAs describing the ABR overlay of Figure 1.

The ABR-LSAs describe the topology of the ABR overlay. Each ABR-LSA includes the identifiers of the originating and neighboring ABRs, and the intra-area shortest path costs from the originating ABR to the neighboring ABRs. When there is more than one connection between two neighboring ABRs, only the lowest cost one is advertised. In our example, the ABR-LSA originated by R6 says that R6 has R3, R4, and R5 as neighbors, and the shortest path costs with these neighbors are 2, 1, and 3, respectively. An ABR obtains this information through the LSDBs of the areas it directly attaches to, namely from their Router-LSAs and Network-LSAs; the information on whether the originating router is an ABR is obtained through the B-bit of Router-LSAs. In our example, router R5 is attached to areas 2 and 4 and, therefore, has access to the LSDBs of these areas. The LSDB of area 4 says that R6 is a neighboring ABR in this area, and is reachable at a cost of 3; likewise, the LSDB of area 2 says that R2 is a neighboring ABR in area 2, and is reachable at a cost of 1.

The Prefix-LSAs describe the address prefixes available at each area. ABRs originate the prefixes of the areas they directly attach to. Each Prefix-LSA includes the identifier of the originating ABR, the advertised prefix and the intra-area shortest path cost from the originating ABR to the prefix. In OSPFv2, this information is obtained from the Router-LSAs and Network-LSAs which, as discussed previously, provide both the topological and addressing information. In OSPFv3, the information is obtained through the combination of Intra-Area-Prefix-LSAs, which describe the prefixes, and Router-LSAs and Network-LSAs, which identify the network elements each prefix is assigned to. In our example, routers R2, R3, and R4, learn, through the LSDB of area 1, that address prefix ap1 is assigned to router R1, and determine that the intra-area shortest path costs to R1 are 4, 1, and 5, respectively.

The ASBR-LSAs describe the ASBRs of each area. ABRs originate ASBR-LSAs to advertise the ASBRs located in areas they directly attach to. Each ASBR-LSA includes the ASBR identifier, and the intra-area shortest path cost from the originating ABR to the ASBR. ABRs know whether a router is an ASBR through the E-bit of Router-LSAs. In our example, routers R5 and R6 analyze the Router-LSAs present in the LSDB of area 4, learn that R8 is an ASBR, and determine that the intra-area shortest path cost from themselves to R8 is 1 and 2, respectively.

With the overlay LSAs introduced above, each ABR can build the graph of the ABR overlay and determine the inter-area shortest path costs from itself to the external destinations, e.g. using Dijkstra's algorithm. ABRs have unrestricted access to all AS routes through this graph; therefore, unlike hierarchical OSPF, globally optimal routing is always assured. In our example, R6 determines through the overlay graph that the shortest path cost from itself to ap1 is 3 (via R4); as seen in previous section, hierarchical OSPF computes a shortest path cost of 5 (via R3), in the same situation.

The information computed by the ABRs must then be injected in the areas they directly attach to. We will reuse the LSAs of the existing OSPF specifications for this purpose. Specifically, the information regarding area-external prefixes is injected into an area through Inter-Area-Prefix-LSAs (OSPFv3) or Network-Summary-LSAs (OSPFv2), and the information regarding ASBRs is injected through Inter-Area-Router-LSAs (OSPFv3) or ASBR-Summary-LSAs (OSPFv2). In this way, our solution is fully transparent to area-internal routers, and only requires modifications (i.e. a new software version) at the ABRs.

As in current OSPF, area-internal routers determine the shortest paths and outgoing ABRs towards each area-external prefix by combining the information injected by the ABRs (through Inter-Area-Prefix-LSAs or Network-Summary-LSAs) with the information provided by their LSDBs. In our example, router R1 learns that the shortest path costs from R2, R3 and R4 (its ABRs) to ap2 are 3, 3, and 2, respectively. Based on the LSDB of area 1, it then determines that the shortest path is via R3 with cost 4. In the case of AS-external prefixes, the information provided by the ABRs must be complemented with the information on the actual prefixes, provided through AS-External-LSAs. In our example, R1 learns about ap3 through the AS-External-LSA injected by R8, which points to R8. Based on the information provided by its ABRs (through Inter-Area-Router-LSAs or ASBR-Summary-LSAs) and the LSDB of area 1, R1 determines that the shortest path to the ASBR that injected ap3 is via R3 with cost 5.

The overlay LSAs can be introduced seamlessly in existing OSPF networks using Opaque-LSAs [1], in case of OSPFv2, and the unknown LSAs feature, in case of OSPFv3 [3]. Since these LSAs must all have AS-flooding scope, the OSPFv2 LSAs must be type-11 Opaque LSAs, and the OSPFv3 LSAs must have the U-bit set, and the (S2, S1) bits with values (1, 0).

The detailed format of the overlay LSAs is shown in Figure 3. The ABR-LSA and the ASBR-LSA identify routers using IPv4 addresses (in the Neighbor Router ID and Destination Router ID fields), as presently done in both OSPFv2 and OSPFv3. The Prefix-LSA advertises IPv4 prefixes (OSPFv2) or IPv6 prefixes (OSPFv3) using the format of the current OSPF specifications. The Metric field in all LSAs include the corresponding intra-area shortest path costs.

The graph of the ABR overlay is kept updated and synchronized at all ABRs through the analysis of the LSDBs. If an ABR senses a modification in one of its LSDBs that impacts the ABR overlay (e.g. failure of a neighboring ABR, change in the shortest path cost to a neighboring ABR, addition or removal of ASBRs or address prefixes), it floods immediately the corresponding overlay LSA. This requires no additions to the existing OSPF specifications. However, there must be a process, similar to the initial LSDB synchronization process, allowing a fast update of the overlay graph when a new ABR joins the network. We introduce the ABR Overlay Request and ABR Overlay Response messages for this purpose. When an ABR detects a new neighboring ABR in one of its LSDBs, it sends to that neighbor an ABR Overlay Request message and the neighbor replies with an ABR Overlay Response message, where each message contains the overlay LSAs currently stored at the sending ABR. If an LSA received through this process is new to the receiving ABR, the ABR disseminates it throughout the AS using the usual flooding procedure. In this way, all ABRs receive fast the LSAs required to update the overlay graph. As in the case of overlay LSAs, this process needs only be implemented at the ABRs and is again fully transparent to area-internal routers.

## IV. CONCLUSIONS

This letter proposes an OSPF extension for multi-area networks that overcomes several limitations of the current OSPF specification. Specifically, with this extension multi-area networks can have arbitrary topologies and globally optimal routing is always achieved. The extension uses a link state inter-area routing protocol, supported on three new LSAs, which describe the topological and addressing information as seen by the overlay of area border routers. Our solution is fully transparent to area-internal routers, and only requires modifying the area border routers.

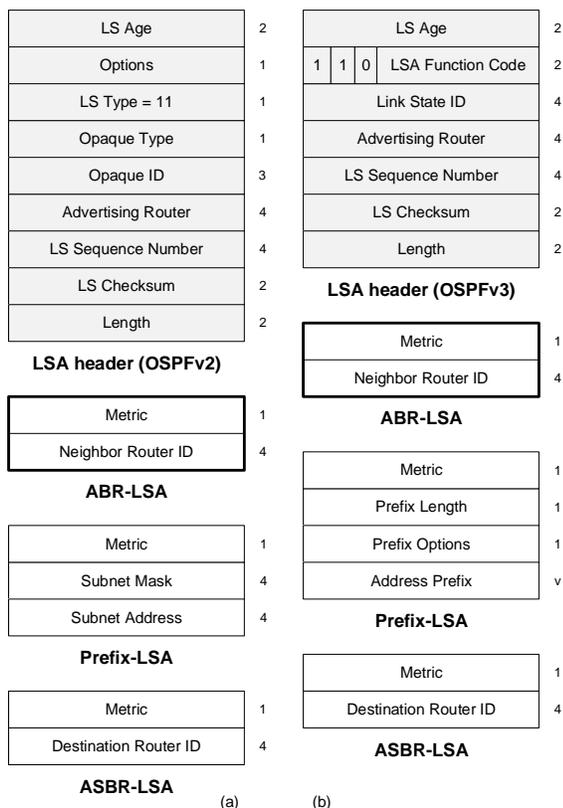

Figure 3: Format of proposed (a) OSPFv2 and (b) OSPFv3 LSAs.